\documentclass[%
 aip,
 amsmath,amssymb,
 reprint,%
]{revtex4-1}

\usepackage{graphicx}
\usepackage{dcolumn}
\usepackage{bm}

\usepackage[utf8]{inputenc}
\usepackage[T1]{fontenc}
\usepackage{mathptmx}
\usepackage{etoolbox}
\usepackage{physics}
\usepackage{caption}
\usepackage{subcaption}
\usepackage{tabularx}
\usepackage{upgreek}

\makeatletter
\def\@email#1#2{%
 \endgroup
 \patchcmd{\titleblock@produce}
  {\frontmatter@RRAPformat}
  {\frontmatter@RRAPformat{\produce@RRAP{*#1\href{mailto:#2}{#2}}}\frontmatter@RRAPformat}
  {}{}
}%
\makeatother
\begin{document}


\title[Low Noise Near-Concentric Optical Cavity Design]{Low Noise Near-Concentric Optical Cavity Design}
\author{Florentin Adam}
 \affiliation{Center for Quantum Technologies, 3 Science Drive 2, Singapore 117543}
\author{Wen Xin Chiew}%
\affiliation{Center for Quantum Technologies, 3 Science Drive 2, Singapore 117543}
\author{Adrian Nugraha Utama}%
\affiliation{Center for Quantum Technologies, 3 Science Drive 2, Singapore 117543}

\author{Christian Kurtsiefer}
 \email{christian.kurtsiefer@gmail.com}
 \affiliation{Center for Quantum Technologies, 3 Science Drive 2, Singapore 117543}
\affiliation{Department of Physics, National University of Singapore, 2 Science Drive 3, Singapore 117542}%

\date{\today}

\begin{abstract}
Near-concentric cavities are excellent tools for enhancing atom--light
interaction as they combine a small mode volume with a large optical access
for atom manipulation. However, they are sensitive to longitudinal and
transverse misalignment. To address this sensitivity, we present a compact
near-concentric optical cavity system with a residual cavity length variation
$\delta L_{C, rms}=0.36(9)$\,\AA. A key part of this system is a cage-like tensegrity
mirror support structure that allows to correct for longitudinal and
transverse misalignment. 
The system is stable enough to allow the use of mirrors with higher cavity finesse
to enhance the atom--light coupling strength in cavity-QED applications.
\end{abstract}


\maketitle

\section*{Introduction}
Establishing strong atom--light interaction is essential for the implementation of quantum networks in atomic systems~\cite{Covey2023,Ballance_2020}.
However, interfacing with atoms can be challenging due to their small cross-section. Commonly used approaches to enhance atom--light interaction involve highly focusing lenses~\cite{Chin2017} or optical resonators~\cite{Hamsen2018}. For the latter, cavity quantum electrodynamics (CQED) has been extensively researched with optical resonators in various configurations~\cite{ReisererRempe2015}, and demonstrated with single atoms since the 1980's ~\cite{Haroche1983,Kimble1999}.
To attain a strong atom--light coupling strength $g$, typically short cavities with a small mode volume and high reflectivity, low loss mirrors are used. This leads to atom--light interactions using cavity mirror spacings ranging from micrometres~\cite{Rempe2020,Reichel_2023} to millimetres~\cite{stamper-kurn2023} with high quality factor $Q$.
Most of the cavities for CQED involve a near-planar geometry, with a mirror separation much smaller than their radius of curvature. This allows for stable mechanical designs, but requires a short distance between the mirror surfaces.

Optical cavities in the near-concentric (NC) regime, where the length of the
cavity is close to the sum of the spherical mirrors' radii of curvature, have
most of the cavity modes strongly focused at the center, leading to a small
effective mode volume~\cite{Durak_2014} and thus a strong atom--light coupling
strength $g$ while providing easy optical access through the relatively large
mirror separation. 

However, NC cavities are challenging to work with compared to planar micro-cavities, as transverse displacement of the mirrors affects the cavity resonance.
As the optical cavity resonance needs to have a well-defined relation to fixed
atomic resonances in CQED applications, the mechanical cavity stability is
critical. In NC cavities, there is the additional requirement for transverse
adjustability and stability, leading to the need for control of three degrees
of freedom for relative mirror positions.

Here, we present a NC configuration of an optical resonator exhibiting low
susceptibility to external mechanical noise and thus stability, while
maintaining adjustability in all necessary degrees of freedom.
The design significantly lowers the mechanical noise compared to our previous
implementation~\cite{Chi_Huan2018}, while allowing us to reach the last
stable point before the concentric point, and hence a stronger atom--light
coupling.

\section*{Cavity design}
To facilitate atom--cavity interaction at a specific transition, in our case the $D_2$ line of $^{87}$Rb
at $\mathrm{\lambda} = 780\,\mathrm{nm}$ with an atomic half-linewidth $\gamma = 2\pi
\times 3.03\,\mathrm{MHz}$, the cavity resonance frequency $\omega_C$ has to match with
the atomic resonance frequency.
To place neutral atoms at the cavity center, we use a magneto-optical
trap\cite{Raab1987} and a dipole trap\cite{GRIMM200095} in an ultra-high-vacuum
(UHV) environment. To retain optical access to the system, and minimize
the size of magnetic coils, a small glass cuvette
with inside dimensions of $25\times25\times150\,\mathrm{mm}$ is
used. The NC cavity system has to fit inside this
relatively small cuvette.

\subsection*{Stability requirement}
\label{sec:stab_req}
To quantify the impact of the shift in cavity resonance and the cavity performance, we introduce a ``noise limit factor'' $\xi$ which normalizes the frequency fluctuation $\delta \omega_C$ due to mechanical noise to the cavity linewidth $2\kappa$. This factor is equivalent to the ratio of mechanical noise (standard deviation $\delta L_{C}$ of the cavity length) to the resonant wavelength, multiplied by the cavity finesse $F$:
\begin{equation}
    \xi = \frac{\delta \omega_C}{2 \kappa} = \frac{\delta L_{C}}{\lambda / 2}\, F .
\label{eq:noise_factor}
\end{equation}
\noindent A noise limit factor $\xi = 1$ indicates that the cavity resonance
fluctuation is equal to its linewidth.
Here, we aim for a target of $\xi=0.15$,
which means that the mechanical noise along the cavity axis contributes to the
cavity linewidth by at most $15\%$.

Both transverse and longitudinal misalignments will affect the reflection
and/or transmission of the NC cavity. Close to the concentric point, transverse positioning noise will dominate the deviation of the
cavity resonance from the atomic transition. For convenience, however, we map
all mechanical effects on the cavity resonance to effective fluctuation of the
cavity length $\delta L_{C}$.

\subsection*{Mirror characteristics}

\begin{figure}[t]
  \centering
  \includegraphics[width=\columnwidth]{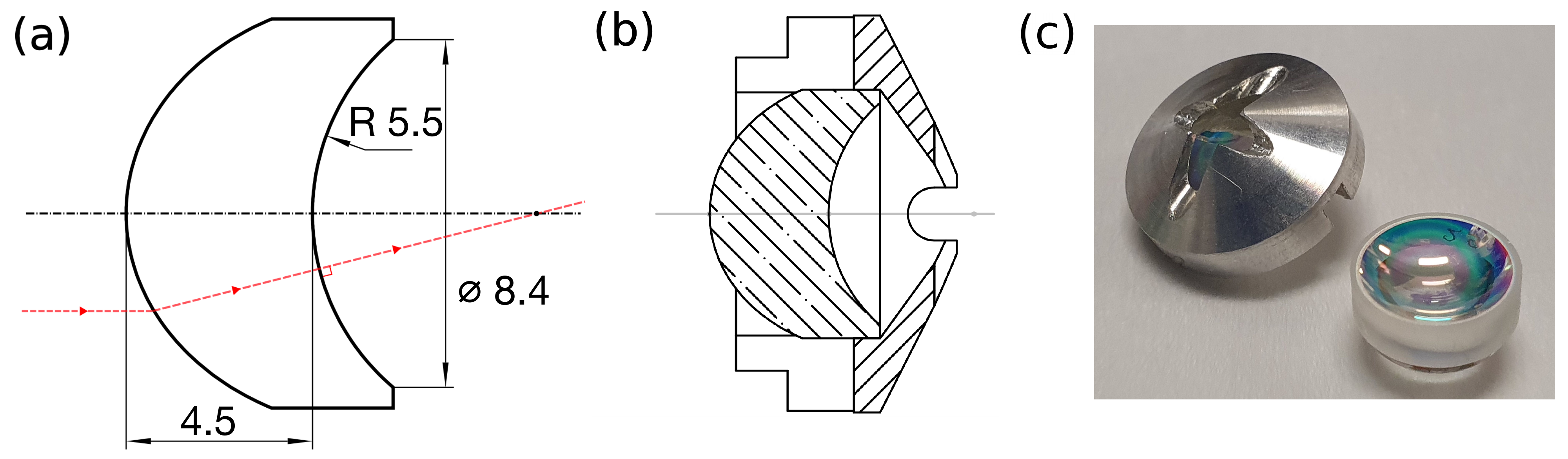}
\caption{Schematic of the cavity mirror: (a) with an example of ray propagation in red through the mirror (units in $\mathrm{mm}$), (b) with the mirror mount, (c) corresponding pictures.}
\label{fig:mirror_spec}
\end{figure}

The cavity mirrors (see Figure \ref{fig:mirror_spec}) have a concave
spherical surface to form the cavity, and a convex ellipsoidal surface,
which provides a straightforward mode matching of a collimated input/output
beam to the highly focused cavity mode~\cite{Durak_2014}. 
The radius of curvature of the concave side is $5.5\,\mathrm{mm}$, resulting
in a cavity length $L_{C} = 2R-d$ in the NC regime,
where $R$ is the radius of curvature and $d$ the (small) critical distance
from the concentric point located at $L_C = 2R =11\,$mm.
The mirrors have a reflectivity of $\mathcal{R} = 99.5\%$ at $\lambda =
780\,\mathrm{nm}$, corresponding to a cavity finesse of $F = 627$.
The target noise limit factor $\xi =0.15$ then corresponds to a cavity length
fluctuation of $\delta L_{C, rms} \approx 0.9\,$\AA.

\subsection*{Cavity support structure}
\label{sec:cav_struct}

To accommodate the cavity mirrors, a structure that allows adjustment of three degrees of freedom of relative mirror positions is required. Moreover, it needs to fit within the constraints of the vacuum system.
Additionally, the structure has to have a low susceptibility to external
noise. To tackle these limiting factors, a tensegrity structure is chosen.

First, the cavity mirrors are fixed with epoxy to metal mirror mounts (see
Figure \ref{fig:mirror_spec} (b)), which protect the mirrors during handling,
and shield them from the possible line-of-sight contamination from the
atomic source. Gaps at the tip of the shield provide optical access for laser
cooling beams to the center of the cavity where the atoms are trapped.
The mirrors in their mounts are fixed to aluminum frames. These are separated
by piezoelectric actuators (PI PICMA P-882.51) with a rectangular profile,
forming the compression members  of the tensegrity structure. Different
orientations of these actuators with respect to the frames were evaluated.

\begin{figure}
    \begin{subfigure}[b]{0.22\textwidth}
        \centering
        \caption{}
        \includegraphics[scale=0.6]{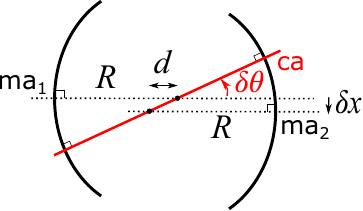}
    \end{subfigure}
    \begin{subfigure}[b]{0.2\textwidth}
        \centering
        \caption{}
        \includegraphics[scale=0.55]{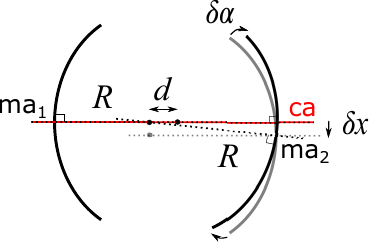}
    \end{subfigure}
    \hfill
    \begin{subfigure}[b]{0.2\textwidth}
        \centering
        \includegraphics[scale=0.08]{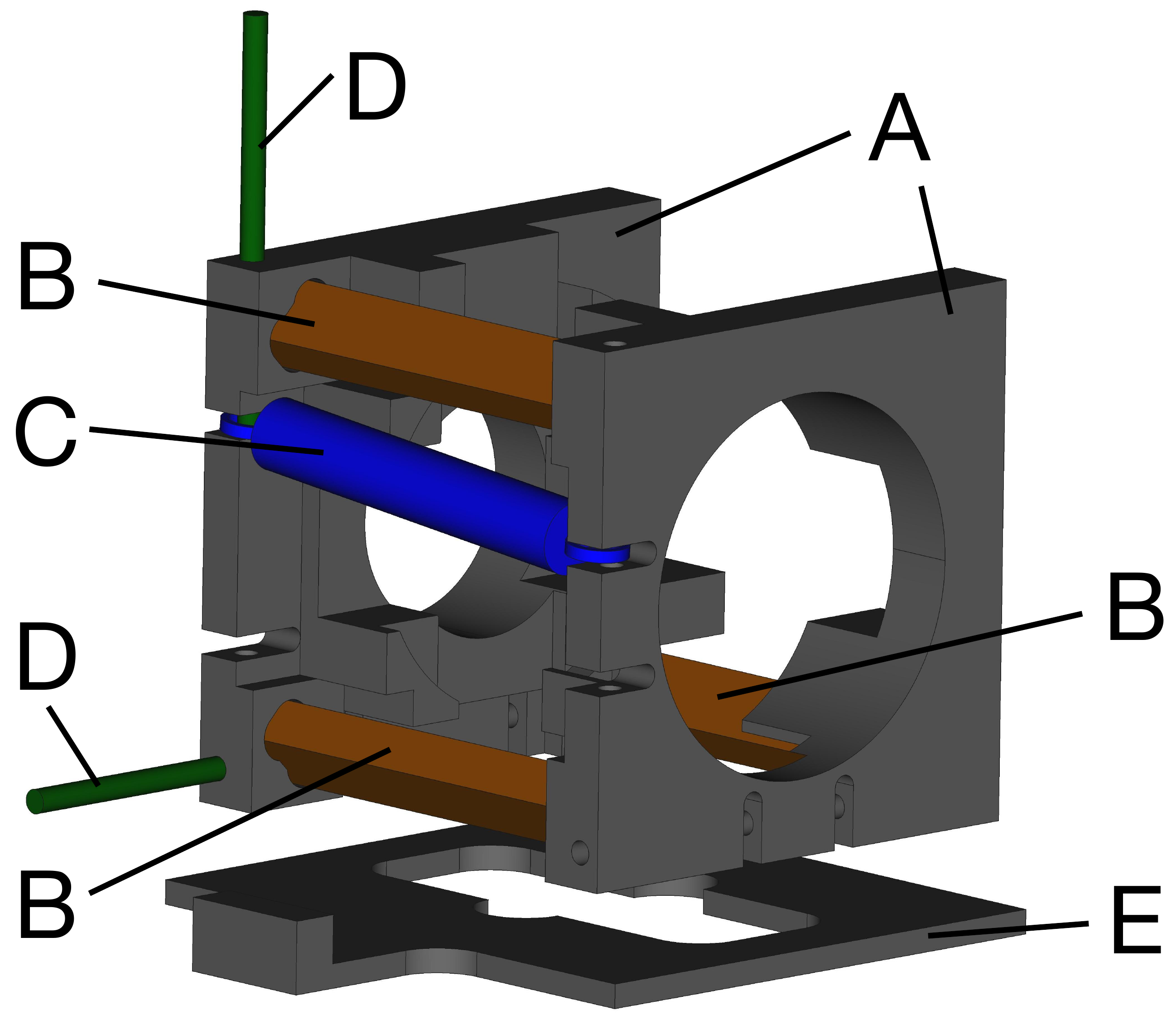}
        \caption{}
    \end{subfigure}
    \begin{subfigure}[b]{0.2\textwidth}
        \centering
        \includegraphics[scale=0.0455]{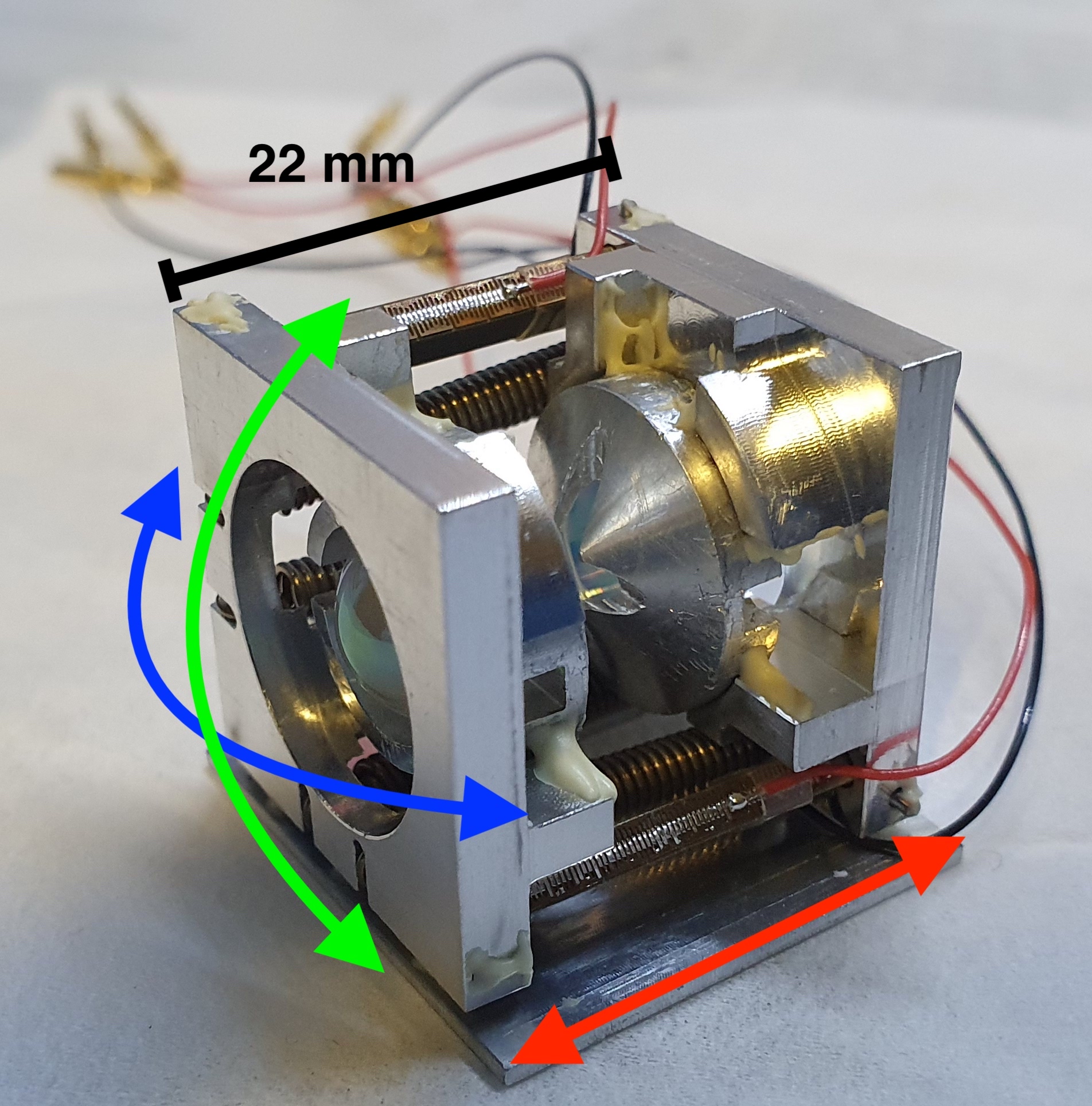}
        \caption{}
    \end{subfigure}
  \caption{(a): Rotation of cavity axis due to transverse misalignment. In a NC cavity of critical distance $d$, a small transverse misalignment $\mathrm{\delta x}$ of the second mirror rotates the cavity axis by an angle $\mathrm{\delta \theta}$. (b): Equivalence of transverse and angular misalignment. A small transverse misalignment $\mathrm{\delta x}$ can be corrected by rotating the mirror by a small angle $\mathrm{\delta \alpha}$. (c) CAD drawing. A: mirror frames. B: actuators. C: spring. D: metal rod. E: transportation plate. (d) Assembled NC cavity.}
  \label{fig:cav_structure}
\end{figure}

To implement the tension members, two alternatives are tested. The first
alternative used a 0.4\,mm thick, laser-cut steel sheet, bent to a clip to
hold the two mirror frames together. The clamp is designed such that only 3
points are in contact with the structure (see Figure \ref{fig:config_clamp}
(a)). The second alternative used helical springs (MISUMI AUT3-20), with
contact points defined by the hooks of the springs (see Figure
\ref{fig:cav_structure} (c) and (d)).

One of the mirror frames is glued (MasterBond EP21TCHT-1) on a stainless steel
plate to avoid applying pressure onto the structure when handling it during
installation in the vacuum chamber, while the other frame can move freely
in order not to constrain the relative tip-tilt movement of the mirrors.

The piezoelectric actuators have a maximal expansion of $15\,\mathrm{\upmu m}$
for an operating voltage of 100\,V, sufficient to compensate
for misalignment after careful pre-alignment. Transverse misalignment can
be corrected via different expansion rates of the three actuators, leading to a
relative tip-tilt motion of the mirrors (see Figure \ref{fig:cav_structure} (a) and (b)).
To allow for independent correction mechanisms for different degrees of
freedom, the tip/tilt corrections $T_{tip}$ and $T_{tilt}$ as well as an
overall cavity length change $\Delta L_{C}$ are combined to the respective actuator voltages $V_{A,B,C}$:
\begin{equation}
    \begin{pmatrix}
        V_A\\
        V_B\\
        V_C
    \end{pmatrix}
    = G
    \begin{pmatrix}
        1 & 1 & 1\\
        1 & -1 & 1\\
        1 & 1 & -1
    \end{pmatrix}
    \begin{pmatrix}
        \Delta L_{C}\\
        T_{tip}\\
        T_{tilt}
    \end{pmatrix}
\end{equation}
\noindent where $G$ is a constant representing the transducer gain. This
allows for both fast changes of the cavity 
length and a slower servo mechanism to maintain transverse cavity
alignment~\cite{Chi_Huan2018}.

\subsection*{Different support structure configurations}

\begin{figure}[!t]
    \begin{subfigure}[b]{0.2\textwidth}
        \centering
        \caption{}
        \includegraphics[scale=0.064]{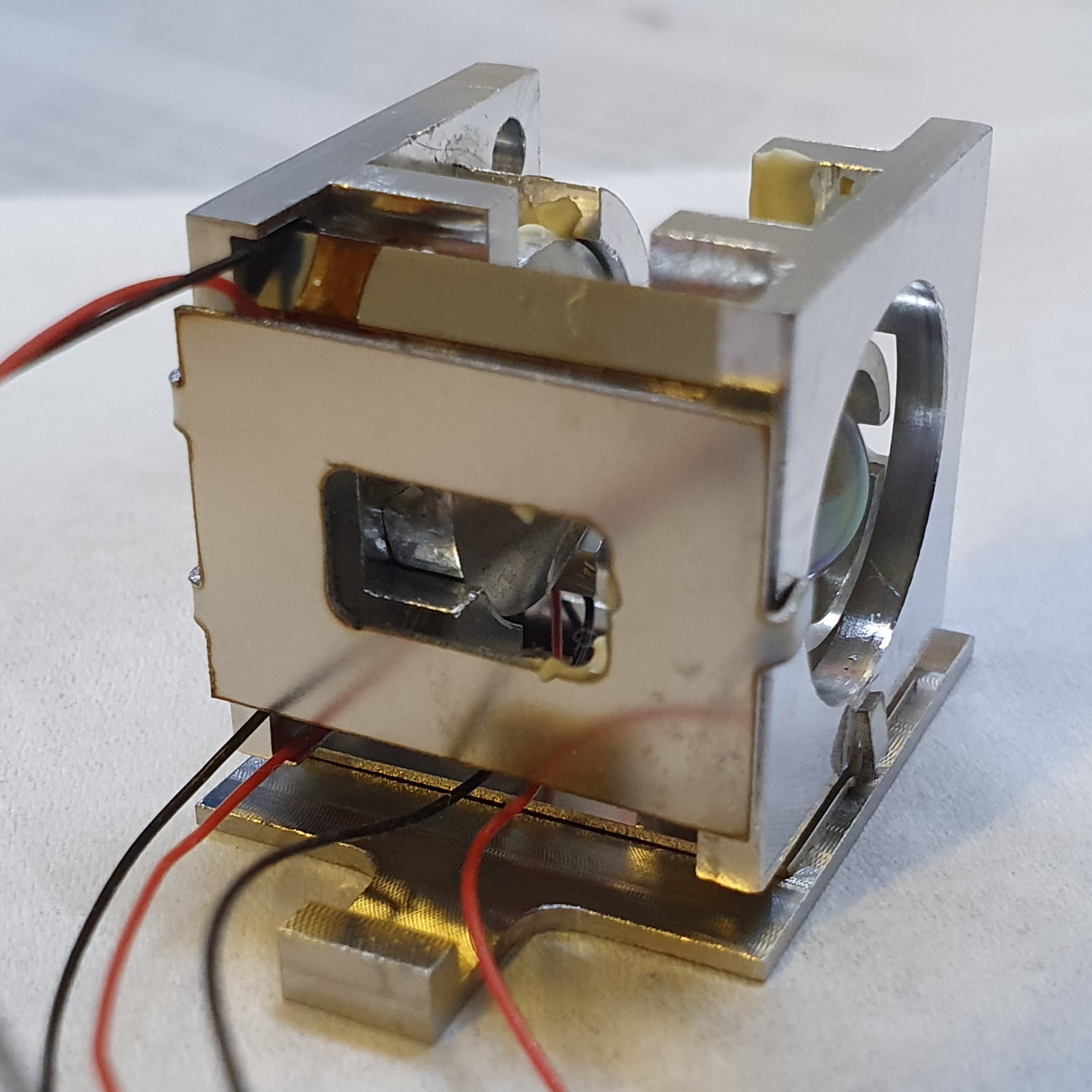}
    \end{subfigure}
    \begin{subfigure}[b]{0.25\textwidth}
        \centering
        \caption{}
        \includegraphics[scale=0.055]{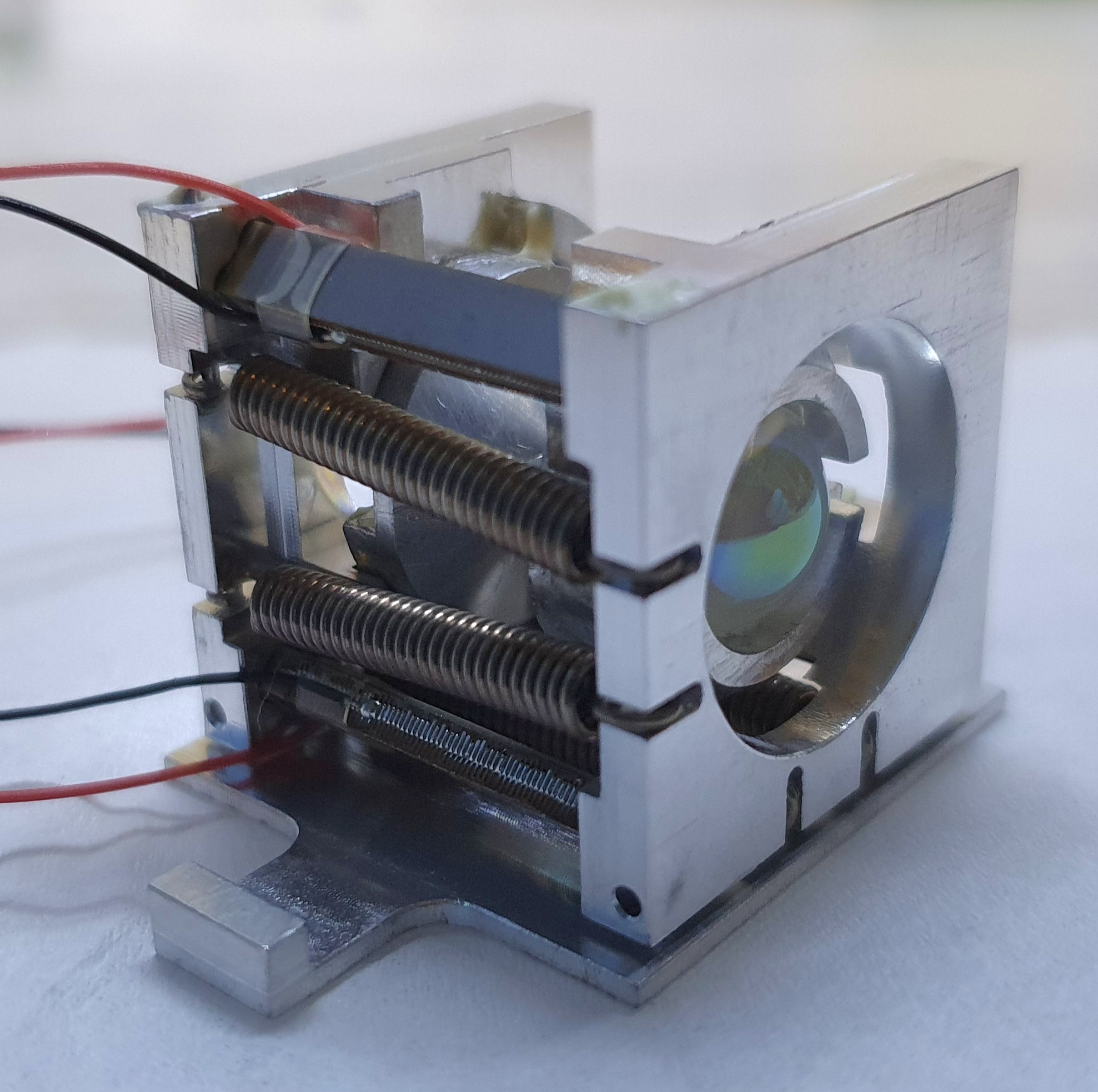}
    \end{subfigure}
    \hfill
    \begin{subfigure}[b]{0.21\textwidth}
        \centering
        \includegraphics[scale=0.072]{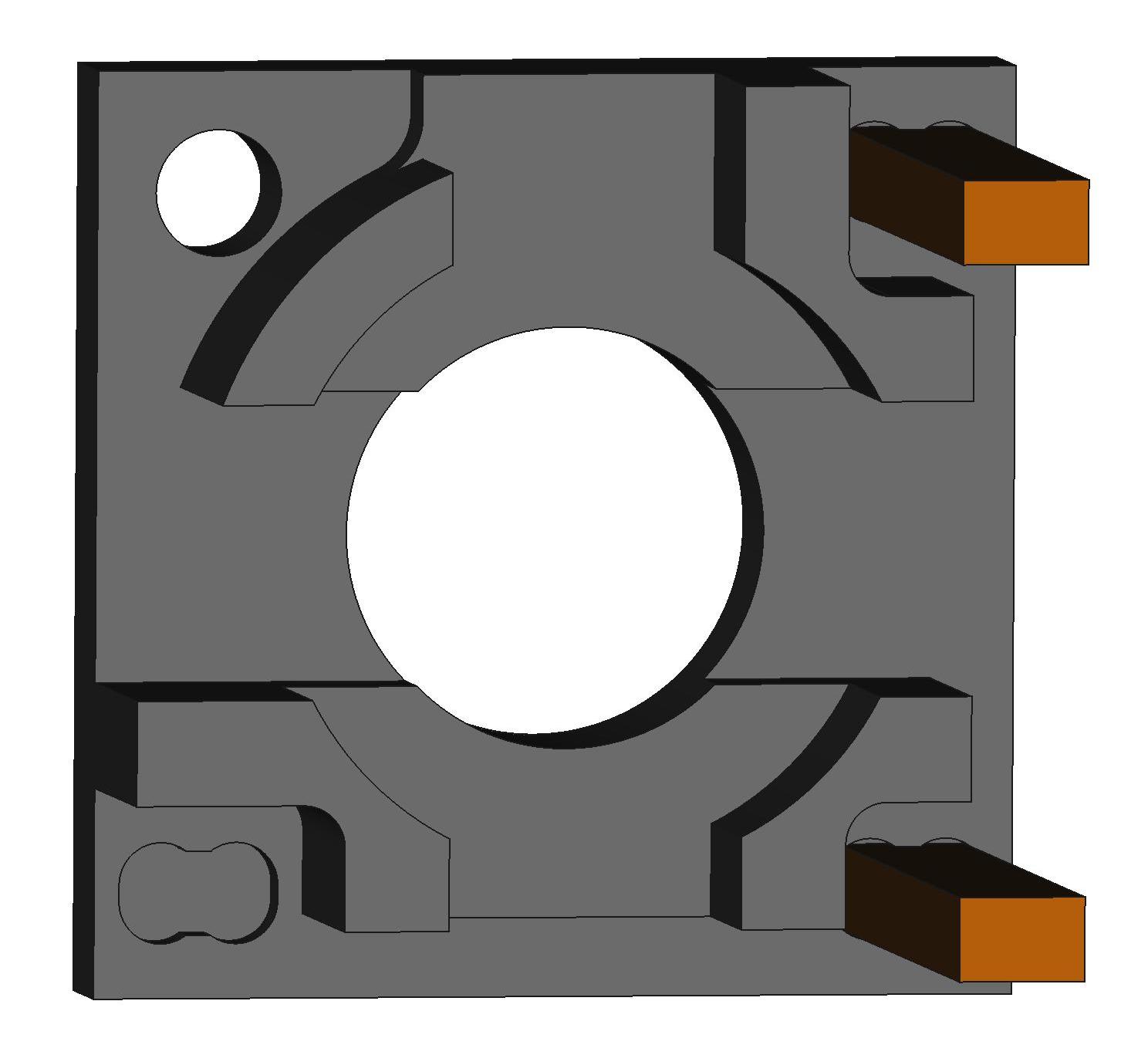}
        \caption{}
    \end{subfigure}
    \begin{subfigure}[b]{0.25\textwidth}
        \centering
        \includegraphics[scale=0.07]{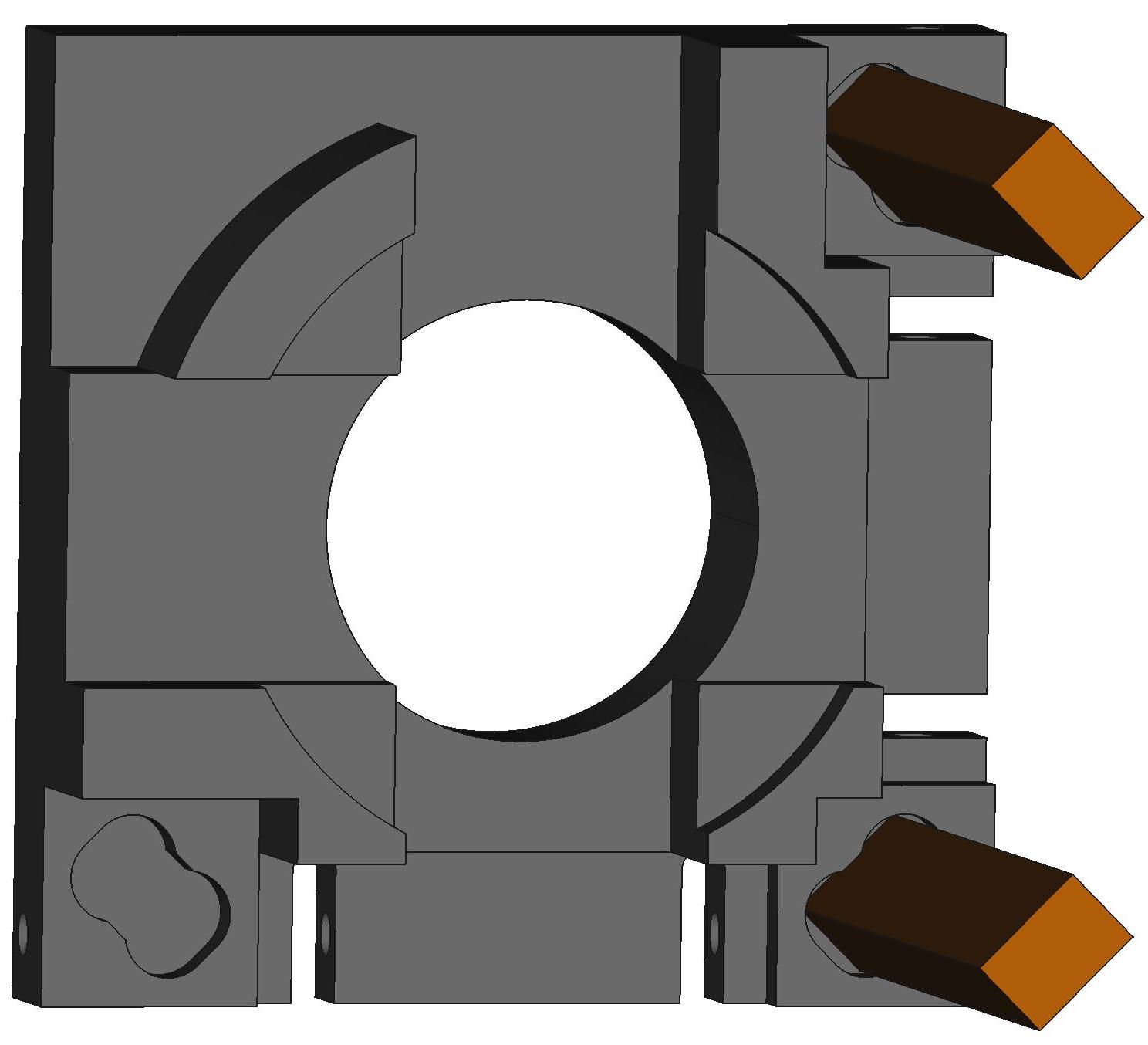}
        \caption{}
    \end{subfigure}
  \caption{(a) Structure using a 3 contact point clamp. (b) Structure using springs. (c) Actuator base parallel to the mirror frame. (d) Actuator base rotated by $45^{\circ}$ from the mirror frame.}
  \label{fig:config_clamp}
\end{figure}

The small expansion range of the piezoelectric actuators limits the correction range
for transverse misalignment. As large transverse displacements can occur in
the cavity structure during the assembly process (epoxy curing, baking of the
vacuum chamber), the structure needs careful pre-alignment. A significant
aspect of this pre-alignment is the static deformation of the compression
structures due to the tension elements, and a consequent transverse
misalignment. Thus, it is necessary to check and reduce the transverse misalignment resulting from the operation of the actuators.

\begin{table}[!b]
\centering
    \caption{Transverse displacement for several structure configurations.}
    \label{tab:cav_disp}
        \begin{tabularx}{\columnwidth}{XX}
            \hline \hline
            Structure configuration & Maximum observed transverse displacement\\
            \hline
            Clamp (Figure \ref{fig:config_clamp} (a)) & \hfill $3.75\,\mathrm{\upmu m}$ \\
            Spring (Figure \ref{fig:config_clamp} (b)) & \hfill $1.25\,\mathrm{\upmu m}$\\
            Parallel base (Figure \ref{fig:config_clamp} (c)) & \hfill $7\,\mathrm{\upmu m}$\\
            $45^{\circ}$ base (Figure \ref{fig:config_clamp} (d)) & \hfill $3.75\,\mathrm{\upmu m}$\\
            \hline \hline
        \end{tabularx}
\end{table}

To evaluate the transverse displacement in the different directions with
regards to individual excitation of each actuator, one mirror frame is fixed
in place with a clamp, and the displacement of the second mirror frame is
measured with a microscope (with a $\times10$ magnifying objective imaged onto a CCD
camera (Point Grey CM3-U3-13S2M-CS)). 
The displacements of each structure configuration is listed in Table
\ref{tab:cav_disp}. The helical springs show a much smaller transverse
misalignment when added to the structure than the flat clip.
We believe this is because the contact points and the static forces of the
clips are not as well defined as the ones from the springs, with their spring
constant tolerance of $\pm10\,\mathrm{\%}$.

Likely due to the actuator's rectangular cross section ($2\, \mathrm{mm}\times
3\, \mathrm{mm}$), an anisotropic bending behavior is observed, with greater
flexural deformation along one axis. Rotating the the actuators with respect
to the frame orientation (see Figure \ref{fig:config_clamp} (c) and (d))
significantly reduced the transverse displacement (see Table
\ref{tab:cav_disp}).

\subsection*{Cavity alignment}

\begin{figure}[!t]
  \centering
  \includegraphics[width=\columnwidth]{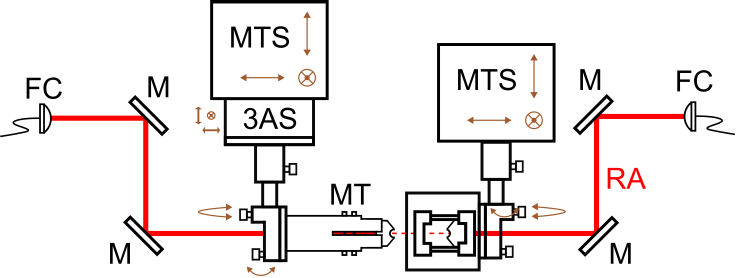}
\caption{Alignment setup of the NC cavity. RA: reference axis, FC: fibre coupler, M: mirror, MT: mechanical tweezer, MTS: manual translation stage, 3AS: three-axis actuator stage.}
\label{fig:setup_align}
\end{figure}

To assemble the near-concentric cavity, a reference axis is established by coupling light between two fiber couplers using four mirrors (see Figure \ref{fig:setup_align}). The cavity system, with the first mirror glued on using a low-outgassing epoxy (MasterBond EP21TCHT-1), is then fixed onto the right tip-tilt stage attached to a three-axis manual translation stage. The combination of the two stages allows the cavity mirror to be freely adjusted along the reference axis. The second mirror is held with a tweezer. The tweezer is also fixed onto the same type of tip-tilt stage as the first mirror, along with a three-axis piezoelectric translation stage with $100\,\upmu\mathrm{m}$ moving range (Piezosystem Jena Tritor 101 CAP) for fine adjustment.

By maximising the coupling of the retro-reflected light from both cavity mirror convex surfaces back into the fibre couplers, each cavity mirror’s axis is aligned along the reference axis (see Figure \ref{fig:setup_align}). The second mirror is then slowly translated towards the first mirror to form a cavity mode. The cavity mode and the transmission spectrum of the cavity are observed with a CCD camera and an amplified photodiode (Thorlabs PDA36A2). Any transverse misalignment is corrected using the three-axis piezoelectric translation stage.

The critical distance $d$ can be estimated from the frequency spacing of the cavity transverse modes. The target critical distance is around $d\approx7.8\,\mathrm{\upmu m}$. This value is chosen as it is around half of the travel distance of the cavity's piezoelectric actuators, which will allow greater tip-tilt tuning as the cavity approaches concentricity.

Once the target critical distance is reached, mirrors are fixed to the frame with a small amount of epoxy, such that additional misalignment during the curing process is minimized. During the initial curing of the epoxy (2 hours), any misalignment is corrected using the three-axis piezoelectric translation stage. When the epoxy is fully cured after another 70 hours, the tweezer is released and removed.

\section*{Cavity stability}

\begin{figure}[!t]
    \includegraphics[width=\columnwidth]{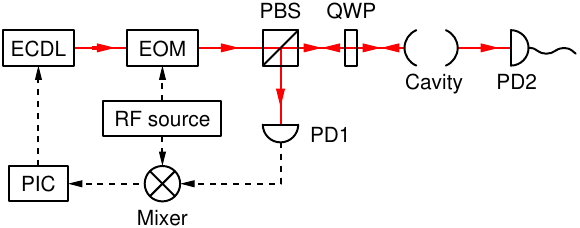}
    \caption{Schematic of the experimental setup. ECDL: external cavity diode laser; EOM: electro-optical modulator; PIC: porportional integral control; PBS: polarized beamsplitter; QWP: quater waveplate; PD: photodiode.}
    \label{fig:setup}
\end{figure}
 
To characterise the susceptibility of the mounted cavity to external noise, the
cavity resonance shift $\delta \omega_C$ is measured with respect to a laser
which is loosely locked to the cavity with a Pound-Drever-Hall (PDH)~\cite{Drever1983} scheme through an integral controller with small gain. 
With this method, fast cavity length changes at high frequencies can be
measured, while ensuring that the error signal remains in the linear regime with respect to the
length change $\delta L_C$, i.e.,  the mapping to frequency detuning stays
injective.

\begin{figure}[!b]
    \includegraphics[width=\columnwidth]{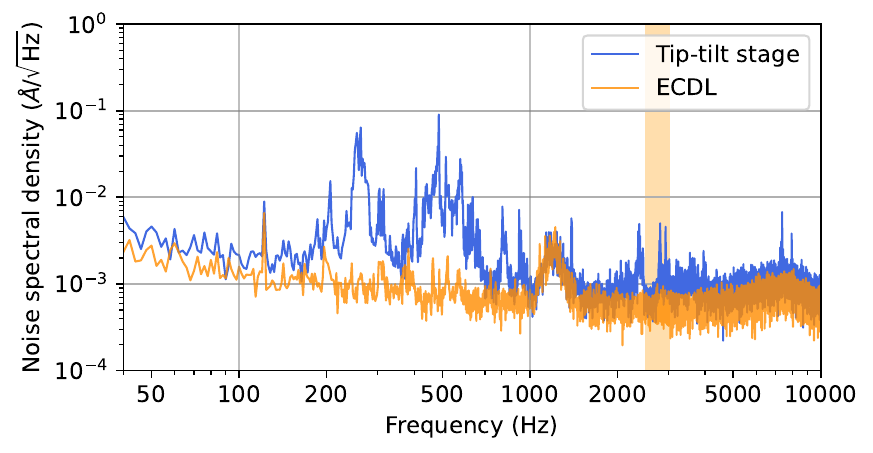}
    \caption{Noise spectral density of the cavity length over an integration
      time of $0.5\,\mathrm{s}$. Total noise is 0.36(9)\,\AA. The shaded region highlights the first resonance of the cavity, centered at $2750\,\mathrm{Hz}$ (see Figure \ref{fig:freq_response}).} 
    \label{fig:noise_density}
\end{figure}

\begin{figure}
  \includegraphics[width=\columnwidth]{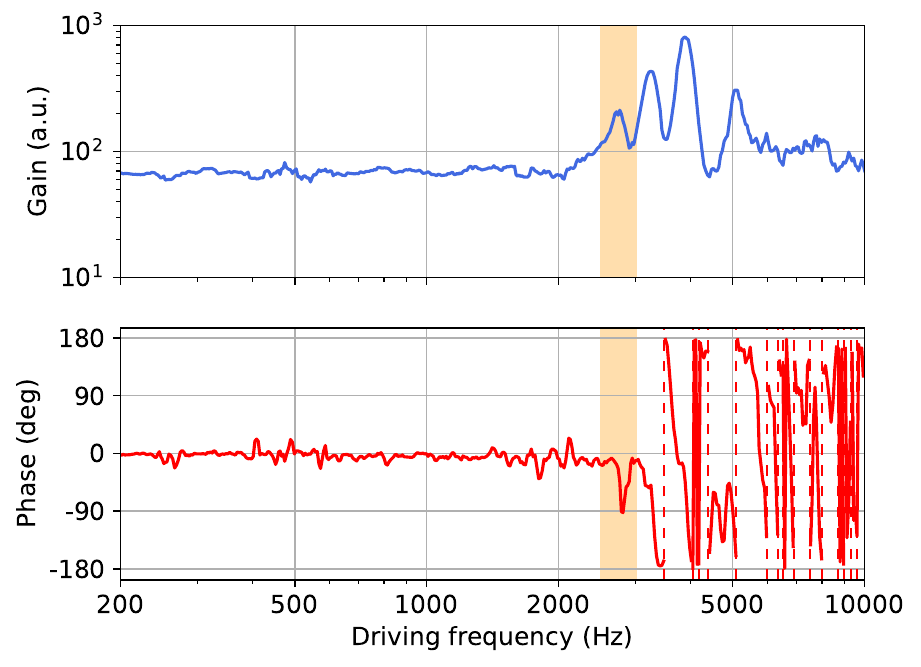}
    \caption{Cavity response to the stimulus sent to the piezoelectric actuators. Shaded region highlights the first resonance of the cavity, centered at $2750\,\mathrm{Hz}$.}
    \label{fig:freq_response}
\end{figure}

The error signal from the PDH scheme is recorded with an oscilloscope and converted to a length change $\delta L_C(t)$. 
The noise spectral density of the recorded length change is shown in
Figure~\ref{fig:noise_density}. The total mechanical noise of the
cavity system over the whole spectrum is $\delta L_{C, rms} = 0.36(9)\,$\AA, or a
corresponding frequency uncertainty of $\delta\omega_C =
1.28(8)\,\mathrm{MHz}$.

The recorded error signal combines both laser noise and cavity noise. To
separate the two contributions, the laser noise is characterised independently
via modulation transfer spectroscopy~\cite{McCarron_2008} using a Rubidium vapour
cell. The corresponding trace is shown in Figure \ref{fig:noise_density} as
well, with an integral total frequency uncertainty of $\delta\omega_{laser} = 0.11(3)\,\mathrm{MHz}$.

More than 70\% of the total noise energy of the laser+cavity system is
contained in a spectral window between 200\,Hz and 2500\,Hz, dominated by
cavity contributions. This noise contribution could be caused by the
susceptibility of transverse vibration modes of the springs to external noise,
ultimately coupling to the cavity length.

To further enhance the system's stability, one can consider a stronger active stabilization scheme, in addition to the existing loose I lock, for the cavity. This becomes feasible with access to a feedback signal for the cavity length, such as through the PDH scheme mentioned earlier. To implement this, we assess the cavity resonance response to an actuator length change stimulus at different frequencies.

A network analyzer (Agilent E5061b) generates this stimulus, which is added to
the actuator's voltage, and picks up the error signal from PDH scheme in the same loose-lock configuration as above. The resulting
Bode diagram of the system response is shown in Figure
~\ref{fig:freq_response}.
A first resonance is observed around $2750\,\mathrm{Hz}$, only contributing 1\% of the total mechanical noise, with a fairly flat
phase response below this resonance. Establishing a phase margin of 60$^\circ$
as the limit of the control for an active stabilisation implementation, an
active length control of the NC cavity system up to a control bandwidth of
$\approx2500\,\mathrm{Hz}$ should be possible, removing the strong broad
contribution up to 2000\,Hz in the observed cavity noise spectrum.
However, in our application, the passive effective cavity length
uncertainty of $\delta L_{C, rms} = 0.36(9)$\,\AA\  is sufficient.

\section*{Conclusion}

We implemented a passively stable compact structure for a near-concentric cavity, with an effective length uncertainty of $0.36(9)\,$\AA. This corresponds to a noise limit factor (see Equation ~\ref{eq:noise_factor}) of $\xi\approx0.05$.
The passive stability permits an increase of the resonator's finesse while
maintaining the cavity's stable relative to its linewidth, such that the
system can operate effectively up to the last stable point of the concentric
cavity. This increase in finesse will give access to even stronger atom--light interaction.
Implementing an active cavity length stabilisation should suppress the susceptibility to external noise even further.

By relying purely on the current susceptibility to external noise of the
cavity system,
we can already enter the strong coupling regime with Rubidium atoms with a
relatively low finesse of 627. With the current cavity parameters, targeting
the $D_2$ cycling transition of $^{87}$Rb, the coupling strength can reach up
to $g = 2\pi \times 17.3\,$MHz, for a cavity decay rate of $\kappa = 2\pi
\times 10.9\,$MHz, leading to a maximum cooperativity of $C = g^2 / 2\kappa
\gamma = 4.5$. This strong coupling will be used to explore atom--light
interaction at the last near-concentric stable point.

In summary, we showed that the near-concentric cavity geometry can provide a viable alternative to near-planar cavity geometries for cavity-QED experiments, offering good optical access to the center of the cavity mode for atomic state preparation in quantum information processing schemes, and a large separation of mirror surfaces from the mode region with a strong field, reducing, e.g., the influence of charges on the mirror surfaces in ion trap configurations.

\section*{Acknowledgements}
This work is supported by the National Research Foundation, Singapore, and
A*STAR under project NRF2021-QEP2-01-P01/W21Qpd0101.

\section*{References}
\nocite{*}
%

\end{document}